\documentclass[aps,prl,twocolumn,showpacs,superscriptaddress,superscriptaddress]{revtex4-1}  
\usepackage{graphicx}  
\usepackage{dcolumn}   
\usepackage{bm}        
\usepackage{amssymb}   
\usepackage{epsfig}
\usepackage{color}

\begin{document}

\widetext

\hspace{5.2in} \mbox{LPS-Pub-CNRS-UMR8502}

\title{Spin Liquid Ground State in the Frustrated Kagome Antiferromagnet MgCu$_3$(OH)$_6$Cl$_2$}
\author{E.~Kermarrec} \affiliation{Laboratoire de Physique des Solides, Universit\'e Paris Sud, UMR CNRS 8502, 91405 Orsay, France}
\author{P.~Mendels} \affiliation{Laboratoire de Physique des Solides, Universit\'e Paris Sud, UMR CNRS 8502, 91405 Orsay, France}
\author{F.~Bert} \affiliation{Laboratoire de Physique des Solides, Universit\'e Paris Sud, UMR CNRS 8502, 91405 Orsay, France}
\author{R.H.~Colman} \affiliation{University College London, Department of Chemistry, 20 Gordon Street, London, WC1H 0AJ, UK}
\author{A.S.~Wills} \affiliation{University College London, Department of Chemistry,  20 Gordon Street, London, WC1H 0AJ, UK}
\author{P.~Strobel} \affiliation{Institut N\'eel, Universit\'e Joseph Fourier, CNRS, 38042 Grenoble, France}
\author{P.~Bonville} \affiliation{Service de Physique de l'\'Etat Condens\'e, CEA-CNRS, CE-Saclay,  91191 Gif-Sur-Yvette, France}
\author{A.~Hillier} \affiliation{ISIS Facility, Rutherford Appleton Laboratory, Chilton, Didcot, Oxon OX11 OQX, UK}
\author{A.~Amato} \affiliation{Laboratory for Muon Spin Spectroscopy, Paul Scherrer Institut, CH-5232 Villigen PSI, Switzerland}

\begin{abstract}
We report $\mu$SR experiments on Mg$_{x}$Cu$_{4-x}$(OH)$_6$Cl$_2$ with $x\sim1$, a new material isostructural to Herbertsmithite exhibiting regular kagome planes of $S=\frac{1}{2}$ ions (Cu$^{2+}$), and therefore a candidate for a spin liquid ground state. We evidence the absence of any magnetic ordering down to 20~mK ($\sim J/10^4$). This report presents a detailed investigation of the spin dynamics on well characterized samples in zero and applied longitudinal fields and a defect-based interpretation is proposed to explain the unconventional dynamics observed in the quantum spin liquid phase.

\end{abstract}

\maketitle



Quantum magnetism in frustrated networks has been acknowledged for a long time as the ideal playground to stabilize new quantum phases. Since the proposal of an RVB state by P.W. Anderson~\cite{Anderson}, advances in theory has led to the emergence of a rich variety of spin liquid phases, including algebraic and gapped spin liquids~\cite{Nature_Review,FrustratedBook}. On the experimental side, Herbertsmithite, ZnCu$_3$(OH)$_6$Cl$_2$, is among the best materials to explore such quantum states, as it combines the highly frustrated two-dimensional kagome lattice and quantum spins $\frac{1}{2}$ of Cu$^{2+}$~\cite{Herbert_MIT}. No magnetic freezing has been detected down to 50~mK $\sim J/4000$~\cite{musr_mendels,Neutron_Helton} and spin-spin correlations are found to be short-ranged as expected for a liquid state~\cite{Neutron_M-deVries}.

Nevertheless, this material deviates from the ``perfect'' image since it has sizable level of Cu-Zn intersite mixing. The replacement of non-magnetic Zn$^{2+}$ ions at the interlayer position by magnetic Cu$^{2+}$ causes almost free $S=\frac{1}{2}$ defects that correspond to $5-10$~\% of the total Cu content, as determined from various techniques (see~\cite{Review_Herbert} for a review). The exact amount of the complementary defect, where a non-magnetic Zn$^{2+}$ impurity induces a spin vacancy in the kagome plane, remains difficult to evaluate quantitatively, due to the similar x-ray scattering factors of Cu and Zn, and the absence of a straightforward magnetic response. As a result, this value has been found to vary with samples and experimental techniques, from 1~\%~\cite{JACS_Tyrel} (from anomalous x-ray scattering) to  5~\%~\cite{RMN_Olariu,detailed} (from NMR).

Very recently, a new series, the ``Mg-paratacamites'' Mg$_x$Cu$_{4-x}$(OH)$_6$Cl$_2$ isostructural to the Zn-based paracatamite, has been successfully synthesized. The similar ionic radii of the diamagnetic Mg$^{2+}$ and Zn$^{2+}$ ions leads to a minimal difference of the crystal structure and hence exchange pathway, resulting in comparable coupling $J \sim 190$~K~\cite{UCL_Mg,Exchange,Singh}. If Cu/Mg mixing is still expected, the difference in their x-ray scattering factors now enables reliable x-ray structure analysis. Chu \textit{et al.}~\cite{MIT_Mg} succeeded in synthesizing materials for $x < 0.75$ where they found a ferromagnetic component in macroscopic susceptibility under $T_\mathrm{C}= 4$~K, attributed to a 3D coupling via interlayer Cu$^{2+}$, similar to the Zn case. A minimal Mg$^{2+}$ substitution within the kagome planes ($\leq 3$ \%) was determined through x-ray diffraction. A different synthesis was recently reported which led to samples with $0.93 \leq x \leq 0.98$~\cite{UCL_Mg}, these correspond to the Herbertsmithite analogues and correspondingly to model quantum kagome antiferromagnets. A small ferromagnetic fraction at T$_c \simeq 4-5$~K was tentatively ascribed to an impurity phase on the basis of susceptibility data.

In this Communication, we first report the $\mu$SR local probe investigation down to 20~mK in the Mg$_x$Cu$_{4-x}$(OH)$_6$Cl$_2$ ``Mg-Herbertsmithites'' with $x\sim1$ which evidences a ground state with no sign of spin freezing, hence a spin liquid character. We also study the sub-Kelvin unconventional spin dynamics of both paratacamites (Mg, Zn) characterized by an $x$-dependent plateau of $1/T_1$. We argue that this relaxation is driven by interlayer Cu$^{2+}$ ions.


The experiments were carried out on ``Mg-paratacamites'' powder samples close to the Herbertsmithite structure, $x$ = 0.84, 0.92, and 1.21, synthesized following a hydrothermal route described elsewhere~\cite{UCL_Mg}, as well as on the Zn Herbertsmithite ($x=1$) sample from~\cite{Defaut_Bert}. The ratios of Cu/Mg and Cu/Zn were determined by ICP-AES (Table \ref{x-ray}). Refinements of x-ray diffraction data for the Mg compounds give separately the Cu ($n$) and Mg ($p$) occupancies of the interlayer and the kagome sites corresponding to the formula (Cu$_{1-p}$M$_p$)$_3$(M$_{1-n}$Cu$_n$)(OH)$_6$Cl$_2$ (M=Mg, Zn). The M total amount is $x=3p-n+1$ (Table \ref{x-ray}). A visible reduction in the quality of the fit, and a corresponding increase in goodness-of-fit statistic ($\chi^2=1.37$ to 1.42 for $x=0.92$), was found when attempting to fix $p=0$ as reported in~\cite{JACS_Tyrel}, evidencing the ability of x-ray diffraction to determine the metal site occupancies of the kagome and triangular sites~\cite{UCL_Mg}. $\mu$SR experiments were performed at the ISIS and PSI facilities in zero and longitudinal applied field configurations down to 20~mK. The DC magnetic susceptibility was also measured on a Quantum Design SQUID magnetometer in the $1.8-300$~K $T$-range and magnetization data were taken at 1.75~K in fields up to 14~T with a Cryogenic Vibrating Sample Magnetometer.

\begin{table}
\caption{\label{x-ray}Chemical composition determined through ICP, x-ray refinements and saturated magnetization for (Cu$_{1-p}$M$_p$)$_3$(M$_{1-n}$Cu$_n$)(OH)$_6$Cl$_2$ where M=Mg, Zn. Each site occupancy is constrained to unity.}
\begin{ruledtabular}
\begin{tabular}{ccccccc}
Element M           & Mg          & Mg          & Zn     & Mg       \\
\hline
ICP         & $x=0.84(1)$ & $x=0.92(1)$ & $x=1.00(7)$  & $x=1.25(3)$ \\
\hline
x-ray       & $x=0.83$\footnotemark[1]    & $x=0.91$\footnotemark[1]    &   -    & $x=1.21(4)$ \\
$p$                 &   0.041(1)      &   0.063(1)      &   -      & 0.12(1)     \\
$n$                 &   0.287(4)      &   0.266(3)      &   -      & 0.15(1)     \\
\hline
Magnetization       &                 &                 &          &          \\
$n$                 &   -             &   0.266(4)      & 0.217(5) & 0.186(4)    \\ 
\end{tabular}
\end{ruledtabular}
\footnotetext[1]{fixed in agreement with ICP analysis.}
\end{table}

\begin{figure}
\includegraphics[width=0.9\columnwidth]{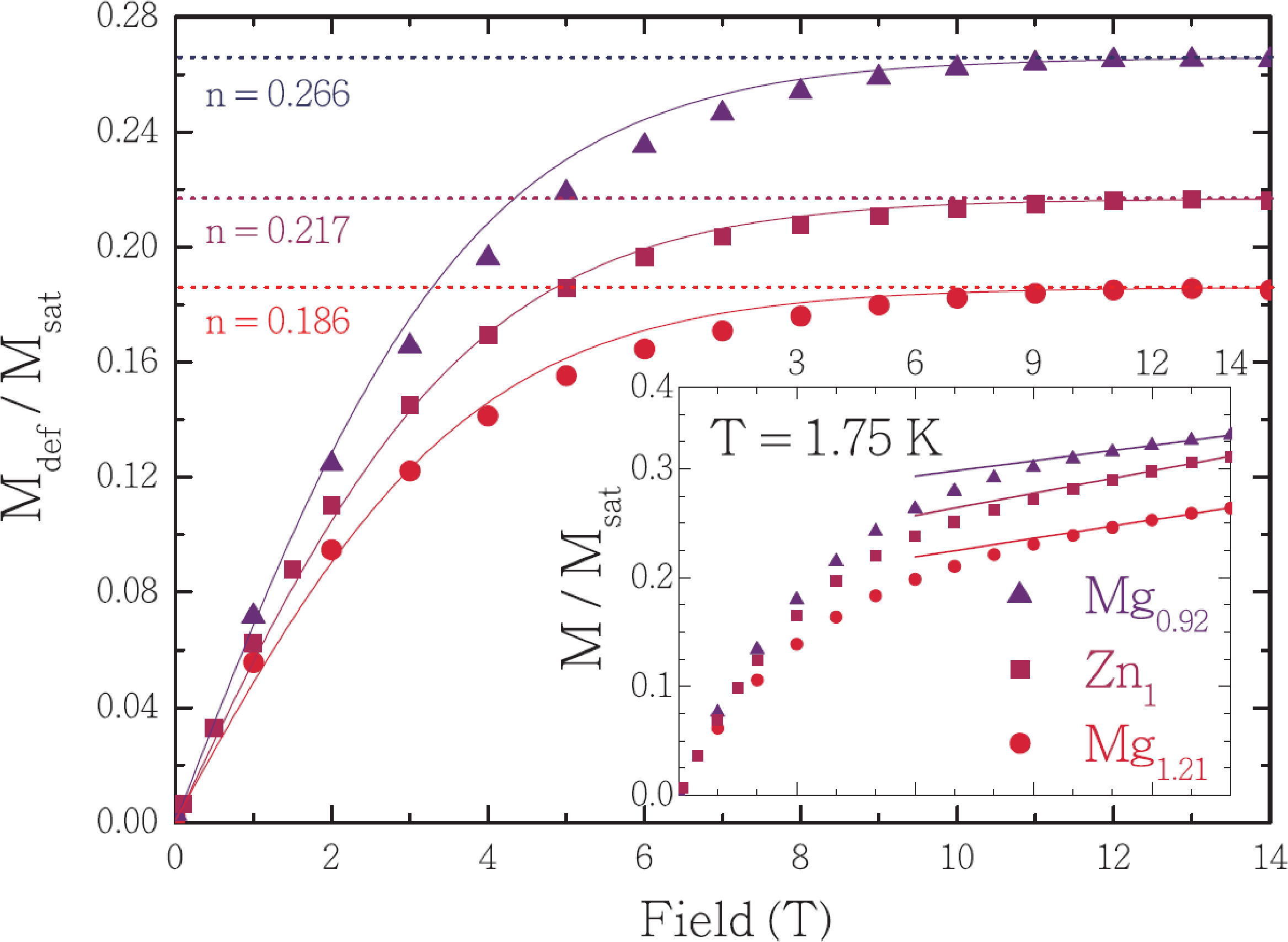}
\caption{\label{Msat} Interlayer Cu$^{2+}$ contribution $M_\mathrm{def}$ extracted from magnetization measurements and normalized by $M_\mathrm{sat} = N_A g \mu_B S = 6143$ emu. The saturation value gives $n$, the amount of interlayer Cu$^{2+}$ per formula unit. The lines reproduce a Brillouin fit with an interacting scale of $\theta = 0.8$~K. Inset: Normalized magnetization at T = 1.75~K up to 14~T. Lines are fits for the linear component.}
\end{figure}

In order to reveal the magnetic response of the interplane defects, we measured the magnetization of the Mg ($x$ = 0.92 and $x$ = 1.21) and Zn ($x$ = 1) samples at low $T=1.75$~K and up to 14~T. Following the analysis in Ref.~\cite{Defaut_Bert,UCL_Mg}, we divide the total magnetization into two contributions, a Brillouin-like component $M_\mathrm{def}$ arising from the weakly coupled interlayer Cu$^{2+}$ and a linear term arising from the strongly coupled Cu$^{2+}$ of the kagome planes (Fig.\ref{Msat}). By subtraction of the latter, we have access to the saturated magnetization, $nM_\mathrm{sat}$, of the interlayer Cu$^{2+}$~\cite{suppl}. Taking $g=2.2$ from Ref.~\cite{DM_Zorko}, this analysis provides an independent determination of $n$, which agrees well with the x-ray refinements for the Mg samples and provides a value for the Zn sample otherwise inaccessible by structural studies (Table \ref{x-ray}). These results confirm quantitatively the existence of the weakly coupled $S=\frac{1}{2}$ interplane defects in the Mg and Zn paratacamites with $x\sim1$.
%
%
%

\begin{figure}
\includegraphics[width=0.9\columnwidth]{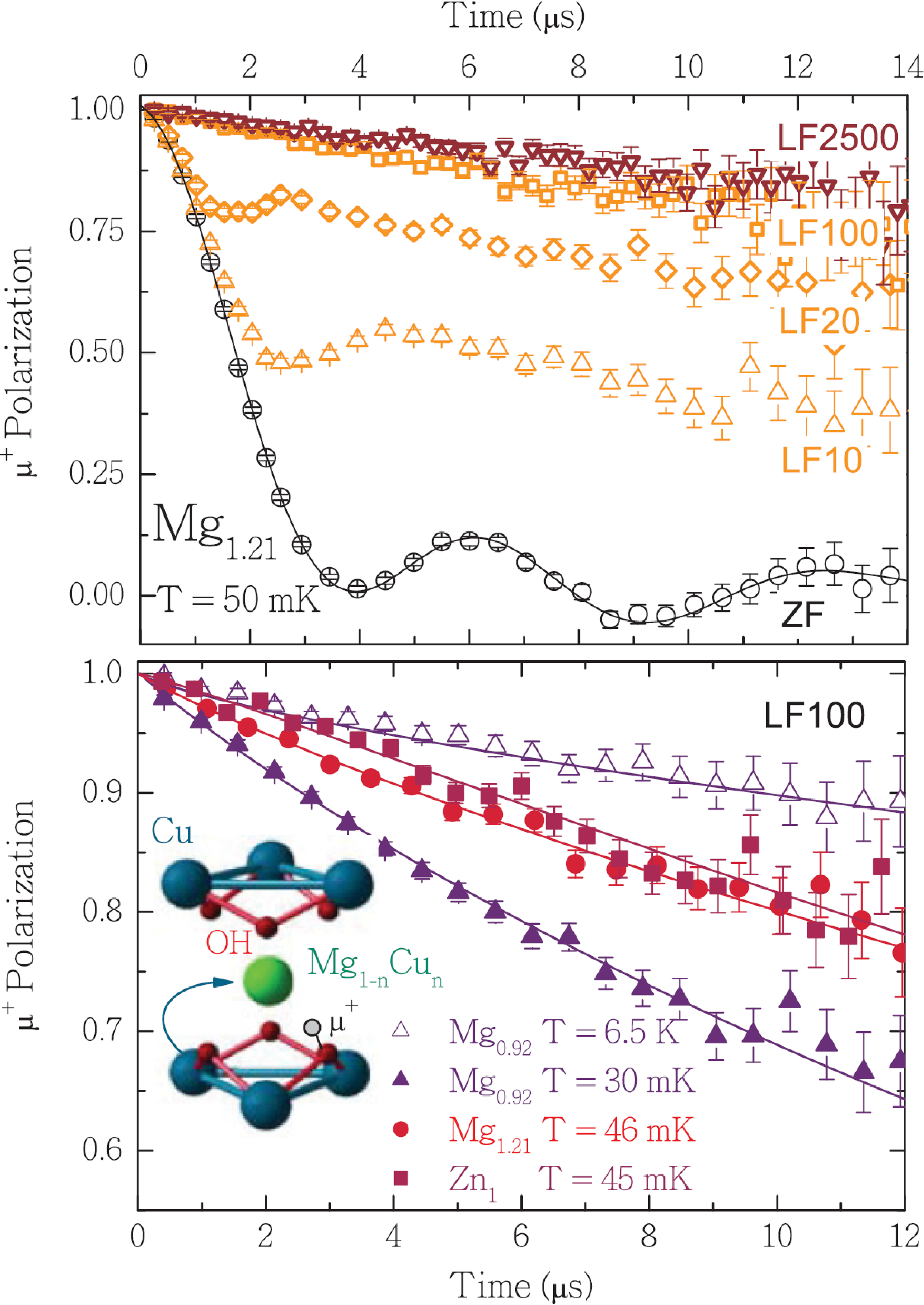}
\caption{\label{decoupling} Upper panel: Polarization in zero field (ZF, black circles) and under longitudinal applied fields (LF) from 10 to 2500 G. The black line is a fit (see text). Bottom panel: The relaxation is faster at low temperatures and when $n$ increases. Lines are stretched exponential fits.}
\end{figure}

We now turn to the local $\mu$SR investigations. $\mu$SR is very sensitive to any small magnetic field (down to $\sim 0.1$~G) and is therefore a powerful tool to detect any frozen moment. The muon is implanted inside the volume and, as a positive charged particle, will stop in the vicinity of a negative environment, \textit{i.e.} either near OH$^-$ or Cl$^-$~\cite{musr_mendels}. The polarizations from 0 (ZF) to 2500 G longitudinal applied field (LF) are reported in Fig.\ref{decoupling}. Our experiments demonstrate a similar magnetic behavior of the Mg and Zn compounds, as expected due to their crystallographic similarities. Following the former work of Ref.~\cite{musr_mendels}, the ZF polarization is fitted on a high statistics run at 50~mK by $P(t) = P_\mathrm{nucl}(t)e^{-\left(\lambda t\right)^{\beta}}$. $P_\mathrm{nucl}(t)$ depends only on static fields from surrounding nuclei, see ~\cite{suppl}, and $\lambda$ stands for a small dynamical relaxation. The well-defined oscillation of $P_\mathrm{nucl}(t)$ is due to the formation of a $\mu$-O-H complex~\cite{ref_POH}, and from the field experienced by the muon, $H_\mathrm{\mu-OH} = 7.8$~G, one estimates a distance of 1.5~{\AA} to hydrogen~\cite{musr_Brewer}. The ZF polarization is found largely unchanging in the extended $T$-range $0.05-20$~K, except for a slight variation of $\lambda$. Therefore, we conclude that there is \textit{no magnetic ordering of the electronic spins} down to $T=20$~mK ($\sim J/10^4$) for $x\sim1$, as in the Zn counterpart.

%
%
%

\begin{figure}
\includegraphics[width=0.9\columnwidth]{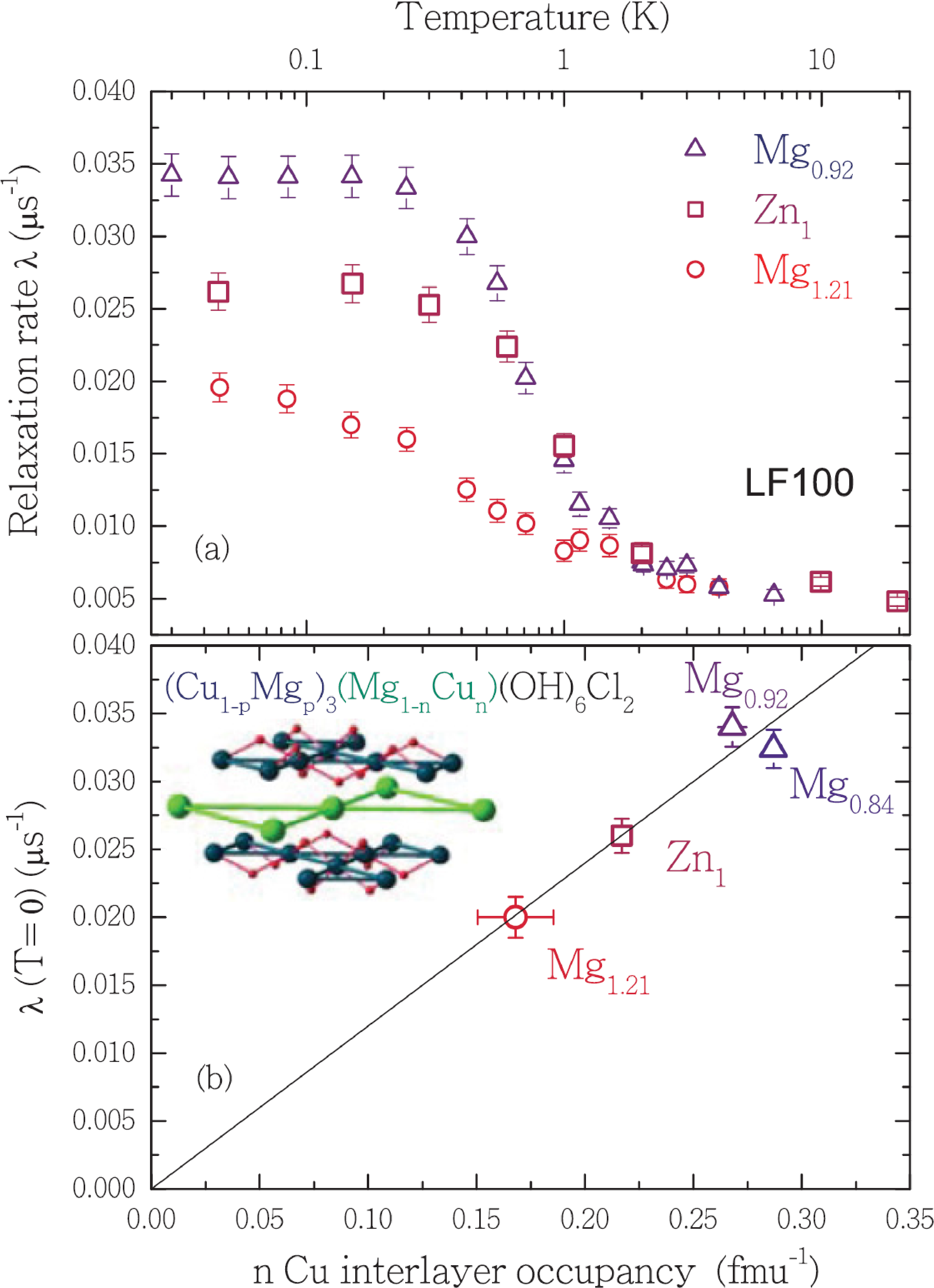}
\caption{\label{relaxation}  (a) $T$-dependence of relaxation rates $\lambda$ for different samples. (b) Plot of the plateau value of $\lambda=1/T_1$ versus $n$. The black line is a linear fit. The error bars on $n$ account for the discrepancy between values derived from $M(H)$ and x-ray characterization.}
\end{figure}

The spin dynamics is revealed by experiments under longitudinal fields. From 10 to 100~G, the static relaxation of nuclear origin is progressively decoupled as expected (Fig.\ref{decoupling}). Above 100~G ($> 10H_\mathrm{\mu-OH}$), $P_\mathrm{nucl}(t)=1$ and a single stretched exponential fit of $P(t)$ yields the relaxation rate $\lambda=1/T_1^\mathrm{\mu}$ of electronic origin. $\lambda = 1/T_1^\mathrm{\mu}$ is linked to the spin autocorrelation function by $1/T_1^\mathrm{\mu} \sim \int_0^{+\infty}\langle \textbf{S}(t) \textbf{S}(0)\rangle\cos(\gamma_\mathrm{\mu}H_\mathrm{LF}t)dt$, where $H_\mathrm{LF}$ is the longitudinal applied magnetic field.  Starting from a high-$T$ value $\sim0.005$ $\mu$s$^{-1}$, $\lambda$ increases upon cooling below 1~K and saturates at lower temperature at a value which differs from sample to sample (Fig.\ref{relaxation}). This increase of $\lambda$ indicates a dynamical slowing down~\footnote{We notice that the shape of the relaxation at low $T$ slightly changes between Mg ($\beta=0.9$) and Zn compounds ($\beta=1.1$).}.

We now argue that the muon is dominantly coupled to the interlayer Cu$^{2+}$ moments from the following experimental observations: (\textit{i}) The low-$T$ muon shift $K^\mathrm{\mu}$ was shown to track the intersite defect susceptibility~\cite{musr_singlecrystal}. This is in fair agreement with recent $^2$D NMR study~\cite{D_NMR} but contrasts with NMR results on $^{17}$O which is strongly coupled to the planes. (\textit{ii}) From Fig.\ref{relaxation} (b) and the inspection of Table \ref{x-ray}, $\lambda$ is found to increase linearly with $n$, the concentration of intersite defects, but is anticorrelated with the level of in plane defects, $p$. Since the distance between Zn sites is large ($d=6.12$~{\AA}), the linearity in $n$ can be easily explained by an ``all or nothing'' model: some muons ($n$) sit next to an interlayer Cu$^{2+}$ defect with a relaxation rate $\lambda_1$ whereas the others ($1-n$) stay far from a defect with a relaxation rate $\lambda_2 \rightarrow 0$. Under the valid condition $\lambda_{1,2}~t \ll 1$, the total polarization becomes $P(t) = n\exp(-\lambda_1 t)+(1-n)\exp(-\lambda_2 t)\sim \exp(-n\lambda_1 t)$, which explains the linear $n$-dependence of $\lambda$. (\textit{iii}) From the scaling of $K^{\mu}$ versus $\chi^\mathrm{bulk}$, one can extract the coupling constant $A_\mathrm{\mu}=0.08$~T/$\mu_B$~\cite{musr_singlecrystal}, consistent with a dipolar interaction. From $^{17}$O NMR $T_1^{17}$, the contribution of the kagome planes Cu$^{2+}$ to the muon relaxation can then be estimated using: $1/T_1^\mathrm{\mu} = 1/T_1^\mathrm{17} \times (\gamma_\mathrm{\mu}A_\mathrm{\mu}/\gamma_\mathrm{n}^{^\mathrm{17}}A_\mathrm{hf}^{^\mathrm{17}})^2 \sim 3\times10^{-5}$ $\mu$s$^{-1}$. This leads at high-$T$ to a value 150 times smaller than the measured one, which therefore requires a different source of relaxation. (\textit{iv}) Finally in the high-$T$ Moriya paramagnetic limit~\cite{Moriya}, $1/T_1^\mathrm{\mu}$ is given by $1/T_1^\mathrm{\mu}=2\gamma_\mathrm{n}^2H_\mathrm{\mu}^2n/\nu$ with $\nu=\sqrt{4J'^2zS(S+1)/3\pi\hbar^2}$ and $H_\mathrm{\mu}=gA_\mathrm{\mu}\sqrt{S(S+1)/3} = 880$~G. From the high-$T$ constant value $1/T_1^\mathrm{\mu} \sim 5\times10^{-3}$~$\mu$s$^{-1}$, an average value $n\sim0.22$ and a number of nearest neighbors $z=6$, one obtains the coupling between an interlayer Cu$^{2+}$ and a kagome Cu$^{2+}$ $J' \sim 3$~K, in rough agreement with the temperature of the magnetic ordering $T_\mathrm{C}=6$~K in the clinoatacamite, Cu$_2$(OH)$_3$Cl, parent compound~\cite{clinoTemp,clinoTemp2,clinoTemp3}.

In this context, the slowing down below 1~K of intersite defects can be attributed either to the strengthening of the correlations with the two nearby kagome planes or to a coupling between two intersite defects mediated by the kagome plane to which they are coupled. The energy scale of 1~K for the interaction between defects was also evidenced in magnetization measurements~\cite{Defaut_Bert}.

\begin{figure}
\includegraphics[width=0.9\columnwidth]{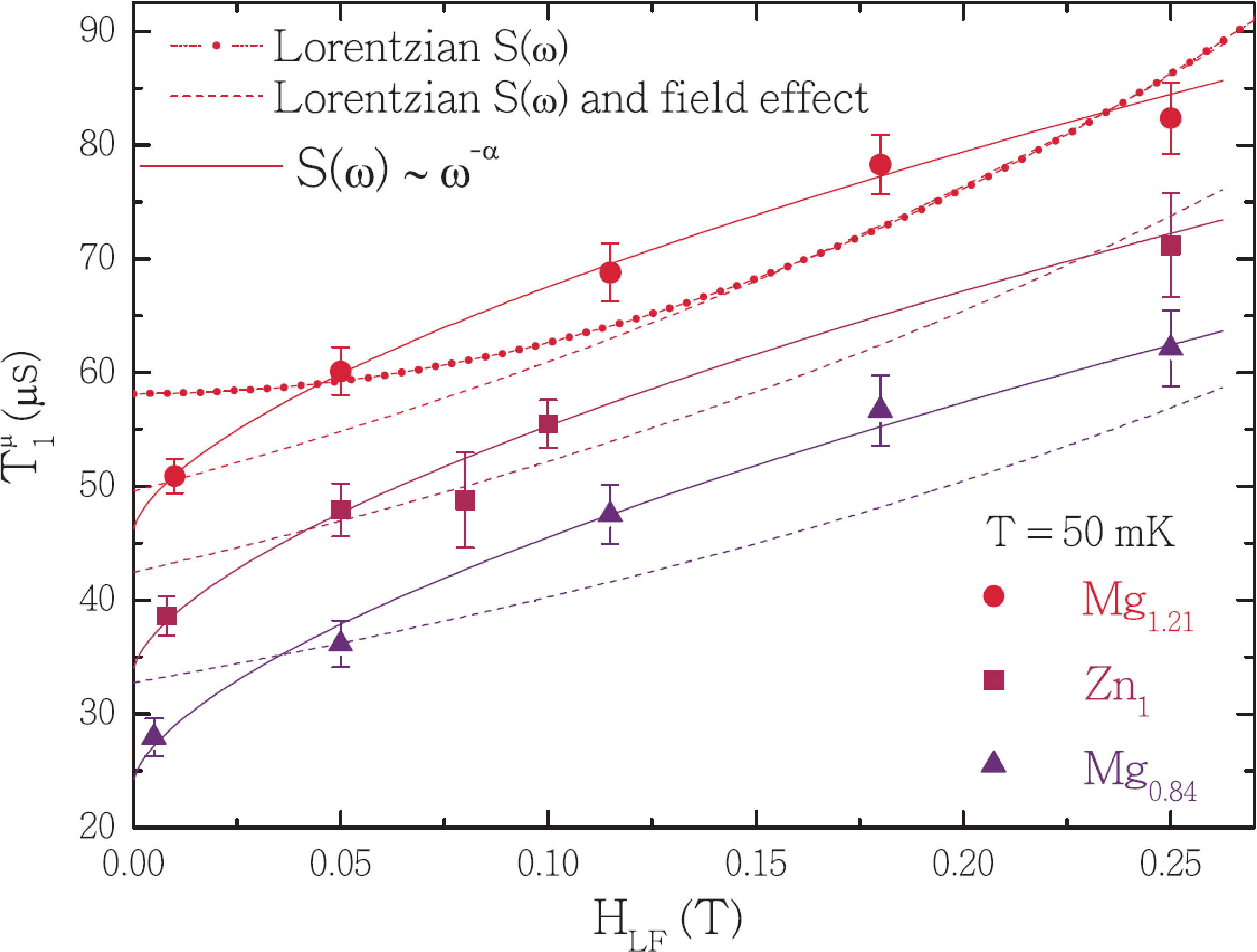}
\caption{\label{Hlinear} The $T_1$ dependence with external field $H_\mathrm{LF}$ underlines the unconventional spin dynamics at $T = 50$~mK. The lines refer to different models (see text).}
\end{figure}

We now discuss the $T \rightarrow 0$ value of the relaxation rate. In order to reveal the dynamics of the correlated regime under 1~K, we probe the excitation spectrum $\tilde{\mathcal{S}}(\omega)$ in the low energy range by applying a longitudinal field $H_\mathrm{LF} = \omega / \gamma_\mathrm{\mu}$, $H_\mathrm{LF} < 0.25$~T. Two scenarios can be considered based on different correlation functions $\mathcal{S}(t)=\langle \textbf{S}(t) \textbf{S}(0)\rangle$ for the interplane spins.\\
\textit{Exponential correlation function $\mathcal{S}(t)=e^{-\nu t}$ combined with field induced polarization of the interlayer defects.}\\
Such correlation leads to the usual Lorentzian spectral density $\tilde{\mathcal{S}}(\omega)$ and
\begin{equation}
\lambda_1 = \frac{2\gamma_\mathrm{\mu}^2H_\mathrm{fluct}^2 \nu}{\nu^2 + \gamma_\mathrm{\mu}^2H_\mathrm{LF}^2}
\label{eq:two}
\end{equation}
where $H_\mathrm{fluct}$ is the fluctuating component of the field at the muon site perpendicular to its initial polarization, $\nu$ is the fluctuation frequency, and $\gamma_\mathrm{\mu}=851.6$~Mrad/s/T is the muon gyromagnetic factor. However, not only is the expected $H_\mathrm{LF}^2$ dependence of $\lambda$ not convincing (Fig.\ref{Hlinear}), but a forced fit results in an unphysical $H_\mathrm{fluct} \sim 20$~G, while the lowest possible dipolar field value is 200~G which corresponds to a maximal distance of the $\mu^{+}$ to intersite Cu$^{2+}$. We therefore propose that the fluctuating field is reduced when the interlayer $S=\frac{1}{2}$ defect starts to be polarized in the external applied field $H_\mathrm{LF}$. In a mean field Brillouin approach, the fluctuating moment is $m^\mathrm{fluct} = \mu_B(1-\tanh(g\mu_B S H_\mathrm{LF}/k_B(T+\theta)))$ where $\theta$ is introduced to account for interactions. The reduced value of $H_\mathrm{fluct}=m^\mathrm{fluct}H_\mathrm{\mu}/\mu_B$ can be injected in Eq.\ref{eq:two} and leads to \textit{only two} shared fit parameters to account for the $T_1^\mathrm{\mu}(H_\mathrm{LF})$ variation for \textit{all} $x$ samples. Although the obtained fits are not perfect, this approach yields $\nu=100\pm 20$~GHz, a typical frequency of fluctuations in a paramagnetic regime, and $\theta=0.7\pm0.2$~K, consistent with both magnetization measurements and the 1~K $T$-scale where the slowing down occurs.\\
%
\textit{Power law correlation function $\mathcal{S}(t)=(1/t)^{1-\alpha}$}\\
The corresponding spectral density $\tilde{\mathcal{S}}(\omega)$ yields a power law relation $T_1^\mathrm{\mu} \propto \omega^\mathrm{\alpha}$. This more exotic approach is at play ($\alpha = 0.35$) in the spin liquid case of the $S=\frac{1}{2}$ antiferromagnetic chain~\cite{1D_Pratt}, where a spinon continuum of excitations is well established. Such a spectral density ($\alpha = 1$) was also invoked for the spin liquid pyrochlore TbTi$_2$O$_7$~\cite{Keren_Tb2Ti2O7,Book_Yaouanc}. A perfect fit of our data can be found with $1/\lambda_1 = T_1^0(x)+A\omega^{0.63}$, which would point to an exotic relaxation channel for the intersite defect. Using the scaling of $\chi''T^\mathrm{\alpha} \sim (T/\omega)^\mathrm{\alpha}\tanh(\omega / \beta T) $ from neutron experiments~\cite{Scaling_Helton}, one gets $1/T_1 \sim k_BT \chi''/\omega \sim \omega^\mathrm{-\alpha}$ by means of the fluctuation dissipation theorem. The reported value $\alpha = 0.66$~\cite{Scaling_Helton} agrees well with that deduced from our data. This points to a common origin and following~\cite{Distribution_J} one could invoke a distribution of couplings between intersite defects as a source of the power law. Another possibility could be that the kagome dynamics drives that of the intersite defects. The inconsistency between the $T$-plateau of the relaxation rate measured at low fields and the $T^{-0.7}$ dependence of $T_1^\mathrm{17}$ at 7~T~\cite{RMN_Olariu} would then require that the field impacts on the kagome plane dynamics.


In conclusion, the spin liquid behavior of the new kagome antiferromagnet Mg$_{x}$Cu$_{4-x}$(OH)$_6$Cl$_2$ for $x>0.84$ is clearly established by our $\mu$SR experiments.  Our data show that the measured interlayer site dynamics differs from those of the kagome plane at 7~T. This calls for a careful inspection of probes which integrate both responses and/or the field impact on relaxation. Whether the exotic character of the $\mu^{+}$ relaxation might also relate to the spin liquid behavior of the kagome lattice is still a matter of speculation. The ability to refine the structure for ``Mg-Herbertsmithite'' in a reliable manner opens the possibility to control the level of defects and to discriminate between the various sources of dynamics at low $T$. Our results open routes for future investigations of the kagome spin liquid ground state in well controlled materials.

This research project has been supported by the European Commission under the 6th Framework Programme Contract nr:RII3-CT-2003-505925 and ANR-09-JCJC-0093-01 grant. We thank C. Baines for assistance at PSI and D. Dragoe for ICP analysis.

\end{document}